\documentclass{elsart}
\usepackage{natbib}
\usepackage{graphicx}
\usepackage{amssymb}
\begin{document}
\begin{frontmatter}
\title{Indirect search for Dark Matter with H.E.S.S.}
 \author[label1]{D. Horns}
\ead{horns@astro.uni-tuebingen.de}
\address[label1]{Institute for Astronomy and Astrophysics, University of T\"ubingen, Sand 1, D-72076 T\"ubingen, Germany}
\author[label2]{for the H.E.S.S. collaboration }
\address[label2]{http://www.mpi-hd.mpg.de/hess}

\begin{abstract}
 Observations of the Galactic center region with the H.E.S.S. telescopes have
 established the existence of a steady, extended source of gamma-ray emission
 coinciding with the position of the super massive black hole Sgr A*.  This is
 a remarkable finding given the expected presence of dense self-annihilating
 Dark Matter in the Galactic center region. The self-annihilation process is
 giving rise to gamma-ray production through
 hadronization including the production of neutral pions
 which decay into gamma-rays but
 also through  (loop-suppressed) annihilation into 
 final states of almost mono-energetic photons.
 We study the observed gamma-ray signal (spectrum and shape) from the
 Galactic center in the context of Dark Matter annihilation and indicate the
 prospects for further indirect Dark matter searches with H.E.S.S.
\end{abstract}
\end{frontmatter}
\section{Introduction}
Observational cosmology has established the existence of non-baryonic
Dark Matter as a clear indication for physics beyond the standard model
of particle physics. Moreover, the overall matter density is dominated
by the contribution of non-baryonic Dark Matter. 
 Despite the good understanding of the cosmological relevance of Dark Matter,
its actual nature remains still unknown. \\
A number of good
 candidate particles 
 for non-baryonic Dark Matter have been suggested in the past: 
 the most intensely studied Dark Matter particle candidate is
 the Lightest Supersymmetric Particle (LSP) which is predicted
 in the supersymmetric extension of the standard model of particle physics
 (Ellis et al. 1984).
 Even though there is currently no experimental evidence for the existence of
 supersymmetric partners of standard model particles, 
 there are convincing theoretical arguments to assume that an additional ``supersymmetry'' exists which e.g. naturally provides a mechanism to break
 the electroweak symmetry. \\
 While supersymmetric particles have not been found in accelerator
 based searches, it is expected that the Large Hadron Collider (LHC)
 at the CERN facility will detect the first evidence for physics beyond
 the standard model. Beyond the LHC, the next generation
 International Linear Collider will (if supersymmetry exists) 
 very likely produce precision 
 data on supersymmetric particles and cross sections (Baltz et al. 2006).\\
 Even though an accelerator based discovery of supersymmetry 
 would give strong reasons
 to assume that Dark Matter is made of supersymmetric particles, it is not
 a proof that this is the case. \\
 Besides accelerator based experiments to create and find Dark Matter particle
 candidates
 in the laboratory, there are two well established approaches to detect
 Dark Matter already present in the Universe: while direct searches aim
 at detecting Dark Matter particles as they traverse laboratory experiments, indirect searches aim at finding particles (as e.g. gamma-rays, neutrinos,
 positrons, and anti-protons) that are produced in the self-annihilation process of Dark Matter particles. 
 Direct and indirect searches for Dark Matter are the only means of 
 tracing Dark Matter in the Universe. While direct Dark Matter searches
 can only probe the Dark Matter density at the position of the solar system, 
 indirect Dark Matter searches offer the advantage of probing Dark Matter 
 in a wide range of Astrophysical environments including 
  the Center of the Galaxy,  Satellite (dwarf) galaxies, 
  black holes environments, nearby Galaxy clusters, as well as the
  possibility of detecting gamma-ray emission from 
  cosmological Dark Matter annihilation (for an overview see 
  e.g. Bertone et al. 2005). \\
  In this contribution, an overview on the observational results of the
  Galactic center region with the H.E.S.S. gamma-ray telescopes and
  their interpretation in the Dark Matter scenario is presented.
\section{Observations of the Galactic Center with H.E.S.S.}
\begin{figure}[h]
  \includegraphics[width=\linewidth]{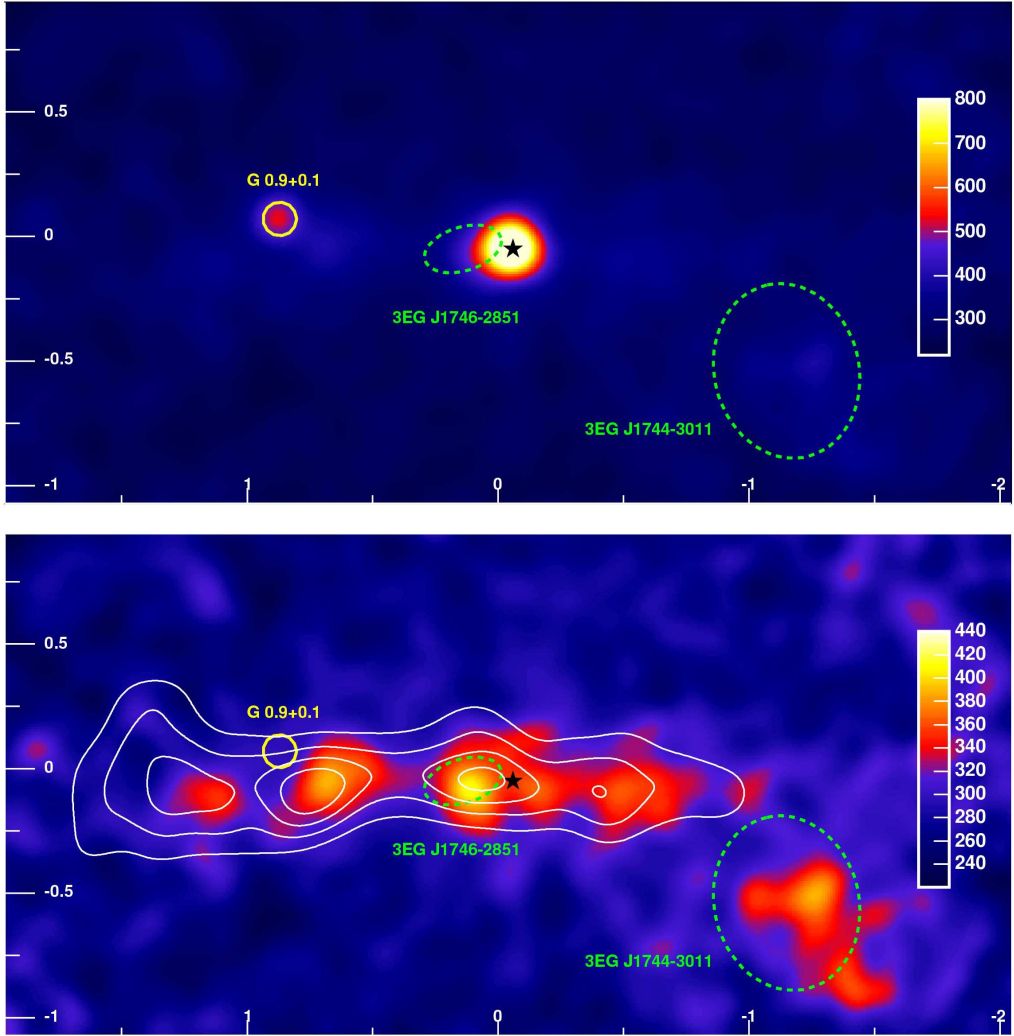}
  \caption{\label{fig:ridge} The upper panel shows the
  smoothed excess map dominated by the signal from the Galactic
  center and from the composite supernova remnant G0.9+0.1. After subtracting
  the contribution from these two sources, an extended emission feature
  is apparent (lower panel) which correlates well with the molecular gas
  density (white contours: smoothed molecular gas density derived from 
  measurements of CS molecular line transitions (Tsuboi et al. 1998)). 
  The star marks the position of Sgr A*, the dashed
  ellipses indicate the 95~\% confidence region of EGRET sources 
  (Mattox et al. 2001) located in the Galactic center region.}
 \end{figure}

\begin{figure}[t]
  \includegraphics[width=\linewidth]{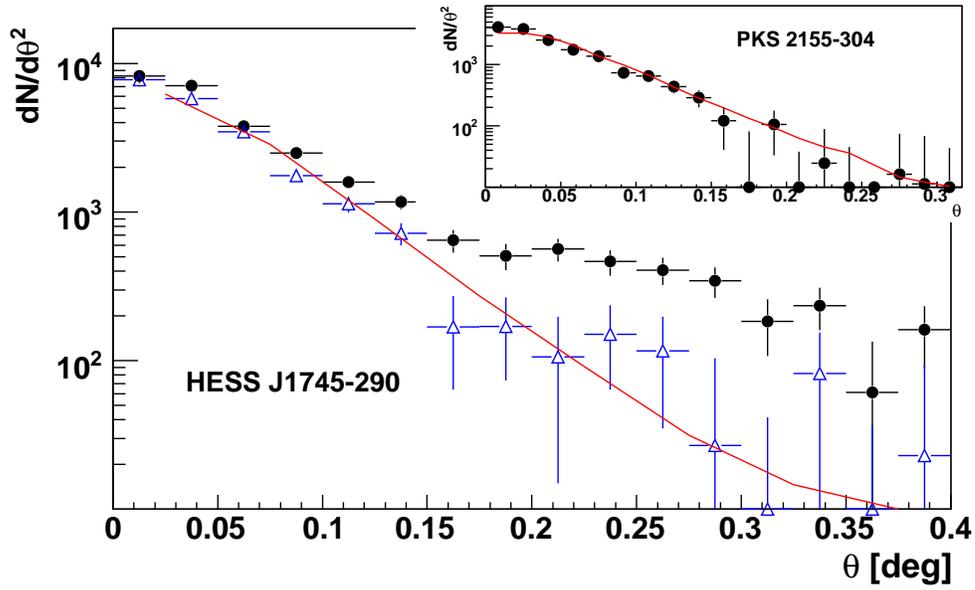}
  \caption{\label{fig:prl} Radial surface brightness 
  profile of the gamma-ray emission
  (number of gamma-ray events per solid angle) 
  from the Galactic center: The solid points indicate the 
  number of gamma-ray events detected from the Galactic center and the
  environment up to an angular separation of $\theta=0.4^\circ$ 
  while the open points show the
  surface brightness after subtracting off the diffuse emission from the
  Galactic ridge. The solid line represents the expected distribution of
  gamma-rays for a point-like source taking into account varying zenith
  angles of the observation. As an example for an observation of a point-like
  source, the signal obtained from the Active Galactic Nucleus PKS~2155-304
  is shown in the inlaid figure.}
 \end{figure}

\begin{figure}[t]
 \includegraphics[width=\linewidth]{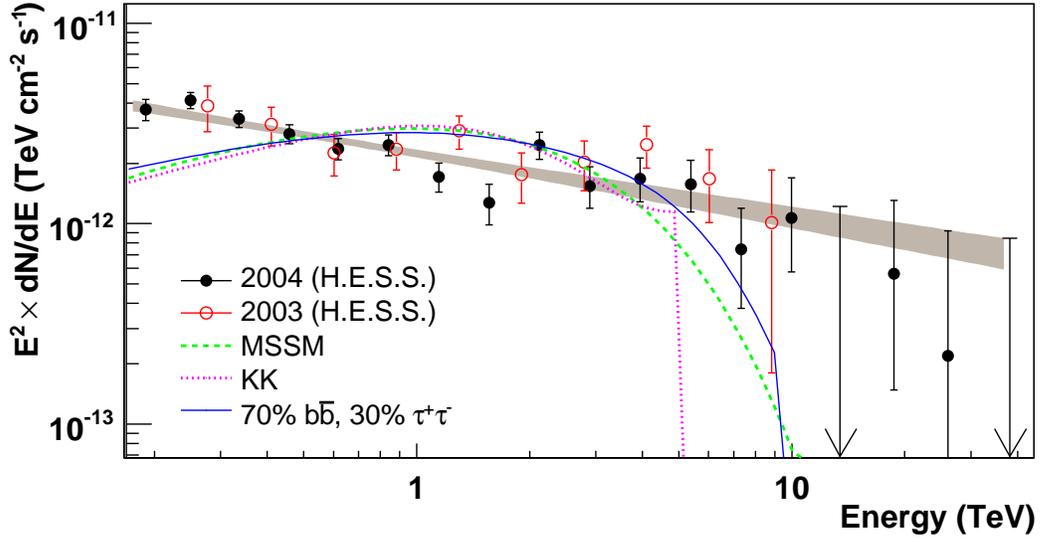}
 \caption{\label{fig:sed} Energy spectrum from the point-like
 source in the Galactic center: the open points are derived from
 the smaller data-set taken in 2003 while the closed points indicate
 the energy spectrum as measured in 2004. The two measurements 
 indicate that the source is steady in time. The shaded region
 shows the power-law fit $dN/dE\propto E^{-\Gamma}$ including
 the statistical uncertainties. The dashed line presents a typical 
 MSSM Dark Matter annihilation spectrum for a best-fit LSP mass 
 $m_\mathrm{LSP}=14~\mathrm{TeV}$. The dotted line gives the
 annihilation spectrum for a Kaluza-Klein type Dark-Matter particle
 with a mass of 5~TeV while the solid line is the annihilation spectrum
 of an LSP with a mass of 10~TeV that annihilates with a specific
 branching ratio of $70~\%$ into $b\bar b$ and $30~\%$ into $\tau^+\tau^-$.
 }
 \end{figure}

 The H.E.S.S. (High Energy Stereoscopic System) 
 experiment is a ground based facility for the
 observation of gamma-rays at energies above 100~GeV. The experiment
 is located in Namibia and consists of an array of four air 
 Cherenkov telescopes positioned at
 the corners of a square with 120~m side length. The first of the four 
 telescopes has been operational since summer 2002, while the full
 array has been taking data since the beginning of 2004
 (Hinton et al. 2004). 
 \\
 The Galactic center region is one of the prime targets of observation
 for the H.E.S.S. telescopes.
 A clear signal from the Galactic center
 was initially obtained after 
 17~hrs of data recorded with the first two telescopes in
 2003 (Aharonian et al. 2004).  
 The detection of gamma-rays from the Galactic
 center have also been reported by the CANGAROO (Tsuchiya et al. 2004), VERITAS
 (Kosack et al. 2003), and the MAGIC collaboration (Albert et al. 2006).\\
 The signal observed with the H.E.S.S. telescopes 
 appeared initially spatially unresolved 
 and the measured energy spectrum 
 covered the range from 300~GeV to 10~TeV. 
 A first comparison of the measured energy spectrum
 with typical self-annihilation spectra 
 indicated that a Dark Matter annihilation scenario could
 reproduce consistently the observations (Horns 2006). 
 However, the estimated mass of the LSP 
 would have to be uncomfortably high: $m_\mathrm{LSP}>10$~TeV. 
\\
 More data (47.8~hrs) were taken in the following year 2004 
 and were used to search for Dark Matter annihilation radiation 
 (Aharonian et al. 2006b).  With the longer observation time, 
 a more detailed study of the source morphology and energy
 spectrum was made possible.
 \\
 As a result of the data analysis, an image of the gamma-ray emission in the central
 100~pc of the Galaxy was obtained (see Fig.~\ref{fig:ridge}). The gamma-ray emission
 is dominated by a point-like source located close to the position of Sgr A* and 
 a gamma-ray source coinciding with the position of a composite supernova remnant G0.9+0.1 
 (Aharonian et al. 2005). After subtracting the contribution of these two sources, a spatially
 extended excess of gamma-rays following the Galactic ridge is revealed (Aharonian et al. 2006a). 
 The surface brightness of the gamma-ray emission is closely correlated with the molecular
 gas density as it is inferred from radio observation of CS molecular transitions (see 
 the lower panel of Fig.~\ref{fig:ridge}). 
 In a Dark Matter annihilation scenario, an extended source of gamma-ray emission from 
 the central part of the Galactic halo is expected. However, the morphology
 of an extended Dark Matter annihilation source is expected to be either
 spherically symmetric or slightly oblate and not to follow the molecular 
 gas density profile. \\
While the extended gamma-ray emission from the Galactic ridge is very likely 
produced by cosmic rays penetrating the clouds (Aharonian et al. 2006a), 
the gamma-ray emission from the point-like
source located at the position of Sgr A* could still be related to Dark Matter annihilation processes.
In order to study the morphology of this source, 
the diffuse emission was now subtracted off the excess map. The resulting
radial surface brightness profile before and after subtracting the diffuse
emission is shown in Fig.~\ref{fig:prl}: the gamma-ray source located at
the Galactic center appears spatially unresolved for the H.E.S.S. telescopes.
The position of the point-source was determined to 
be $\alpha=17^\mathrm{h}45^\mathrm{m}39.44^\mathrm{s}\pm0.6^\mathrm{s}$ and
$\delta=-29^\circ00'30.3''\pm 9.7''$ in equatorial coordinates quoting
statistical errors only. This is within $7''\pm 14''_\mathrm{stat}\pm 28''_\mathrm{syst}$ from the position of Sgr~A*. \\
Based upon the good agreement of the observed radial profile with the expectation of
the point-spread function,
an upper limit on the extension of the source of $1.2'$ at the
95~\% confidence level was derived (Aharonian et al. 2006b). 
\\
In a similar approach as used initially by Horns (2006),
the observed radial profile is compared with the expectation of
a Dark Matter annihilation signal. This is done by folding the
point spread function of the instrument with the
integrated gamma-ray emissivity from Dark Matter annihilation along the
line of sight. 
The gamma-ray emissivity 
is proportional to $\rho^2_\mathrm{LSP}(r)$, with $\rho_\mathrm{LSP}(r)$ 
indicating the radial Dark Matter density profile. Under the assumption
of a power-law type density profile ($\rho_\mathrm{LSP}\propto r^{-\alpha}$),
the observations can be used to obtain a constraint on $\alpha>1.2$ (95~\%
c.l.).
\\
Given that the emission is consistent with a point-like source, the gamma-ray
spectrum from the excess events within $0.1^\circ$ of the position
of the source were used to accumulate an energy spectrum. The
contamination of the diffuse emission within $0.1^\circ$ is
expected to be a comparably small fraction of 16~\% of the total events from
the point source. The resulting energy spectrum is shown in Fig.~\ref{fig:sed}
together with the energy spectrum from the data obtained in 2003. \\ 
The energy spectrum is well fit with a simple power-law: 
$dN/dE=N_0 (E/1~\mathrm{TeV})^{-\Gamma}$ with
$\Gamma=2.25\pm0.04\mathrm{(stat.)}\pm0.10\mathrm{(syst.)}$. The data obtained in 2003 and 2004
are in good agreement with each other indicating that neither the flux nor the shape of the
spectrum have changed with  time.
The integrated flux above 1~TeV is measured to be
$(1.87\pm 0.10\mathrm{(stat.)}\pm 0.30\mathrm{(syst.)})\times
10^{-12}$~cm$^{-2}$s$^{-1}$. An exponential cut-off below 9~TeV 
in the energy spectrum is excluded at the 95~\% c.l. \\
A search for variability on different time-scales has been carried
out on the available data from 2003 and 2004. No indications for
variability have been found. 
\section{Dark Matter interpretation for the Galactic center signal}
Under the assumption that all the gamma-ray emission observed
from the Galactic center source is produced via Dark Matter annihilation, 
it is interesting to compare  
the shape of the observed gamma-ray spectrum with the one expected
from Dark Matter annihilation.\\
It is obvious, that
the observation of a gamma-ray signal up to 10~TeV would require a sufficiently
massive LSP. Furthermore, the observed energy spectrum appears 
to follow a smooth power-law without indications for curvature. 
A curved energy spectrum is however expected for the
most commonly considered supersymmetric models like Minimal Supersymmetric
extension of the Standard Model (MSSM) (see e.g. Ellis et al. 2002), 
Anomalously mediated Symmetry Breaking (AMSB) (see e.g. Profumo \& Ullio 2004),
or Kaluza-Klein (KK) (see e.g. Servant \& Tait) 
scenarios with extra-dimensions. \\
 The fact that the required mass of the LSP is larger than 10~TeV
leads to an additional hard radiative component as suggested
by Bergstr\"om et al. (2006). 
Additionally, some fine-tuning of the model-specific
branching ratios for the annihilation processes can be used to 
produce a harder energy spectrum (Profumo 2006). A number of different
Dark Matter annihilation spectra including a KK spectrum are shown
in Fig.~\ref{fig:sed} together with the measurements. It is evident, that
the data and the considered models do not agree.\\
In a different approach, we assume that the observed gamma-ray emission
is predominantly emitted by  a possible ``conventional'' gamma-ray source which produces
a typical power-law type energy spectrum. Now, we can vary the mass and 
the annihilation rate of the LSP to evaluate the maximum annihilation rate for a
given mass that does not violate the measured energy spectrum.
Following this approach,
upper limits on the annihilation rate can be obtained.  The upper limits
derived through this method are not severely constraining the annihilation cross section 
because of the large
uncertainties on the Dark Matter halo density which can be as large as three
orders of magnitude.
\section{Summary and Outlook}
 The indirect search for Dark Matter using gamma-rays offers a unique 
 way of probing the Dark Matter density in the Universe. The new
 ground based gamma-ray telescopes like H.E.S.S. have in principle achieved sufficient sensitivity
 to detect gamma-ray emission from the annihilation of Dark Matter. The
 drawback of this technique is the difficulty to disentangle the contribution
of Dark Matter annihilation radiation from the emission of conventional gamma-ray sources. 
 The example of the gamma-ray source in the Galactic center demonstrates clearly these difficulties.
 \\
 Whatever the origin of the gamma-ray emission from the Galactic center 
 is, observations at energies below 100~GeV and above 10~TeV
 are  of crucial importance in order to determine the dominant gamma-ray production mechanism.\\
 The H.E.S.S. installation is currently extended (``Phase II'') to include a 
 Large Cherenkov Telescope (LCT) which will extend the accessible energy range of H.E.S.S. to 
 energies as low as 20~GeV.
 This energy range is also accessible with the GLAST satellite. In a few years from
 now, the gamma-ray spectrum from the Galactic center source is expected to be measured 
 from energies below 1~GeV up to energies well beyond 10~TeV, 
 covering a total of four decades in energy.
At the same time, the accelerator experiments of the LHC will provide
the first indications and/or constraints on supersymmetric particles.\\
The indirect search for gamma-ray emission from Dark Matter annihilation using H.E.S.S. Phase II, GLAST,
as well as the coming generation of accelerator experiments will provide us with a clear and detailed view on
the elusive Dark Matter particle.
\section*{Acknowledgments}
\begin{minipage}{\linewidth}
The support of the Namibian authorities and of the University of Namibia
in facilitating the construction and operation of H.E.S.S. is gratefully
acknowledged, as is the support by the German Ministry for Education and
Research (BMBF), the Max Planck Society, the French Ministry for Research, the CNRS-IN2P3 and the Astroparticle Interdisciplinary Programme of the
CNRS, the U.K. Particle Physics and Astronomy Research Council (PPARC),
the IPNP of the Charles University, the South African Department of
Science and Technology and National Research Foundation, and by the
University of Namibia. We appreciate the excellent work of the technical
support staff in Berlin, Durham, Hamburg, Heidelberg, Palaiseau, Paris,
Saclay, and in Namibia in the construction and operation of the
equipment.
\end{minipage}


\end{document}